# Large-Scale Security Analysis of Real-World Backend Deployments Speaking IoT-Focused Protocols

Carlotta Tagliaro
TU Wien
carlotta@seclab.wien

Martina Komsic
TU Wien
e1634065@student.tuwien.ac.at

Andrea Continella
University of Twente
a.continella@utwente.nl

Kevin Borgolte
Ruhr University Bochum
kevin.borgolte@rub.de

Martina Lindorfer
TU Wien
martina@seclab.wien

## Abstract

Internet-of-Things (IoT) devices, ranging from smart home assistants to health devices, are pervasive: Forecasts estimate their number to reach 29 billion by 2030. Understanding the security of their machine-to-machine communication is crucial. Prior work focused on identifying devices' vulnerabilities or proposed protocol-specific solutions. Instead, we investigate the security of backends speaking IoT protocols, that is, the backbone of the IoT ecosystem.

We focus on three real-world protocols for our large-scale analysis: MQTT, CoAP, and XMPP. We gather a dataset of over 337,000 backends, augment it with geographical and provider data, and perform non-invasive active measurements to investigate three major security threats: information leakage, weak authentication, and denial of service. Our results provide quantitative evidence of a problematic immaturity in the IoT ecosystem. Among other issues, we find that 9.44% backends expose information, 30.38% CoAP-speaking backends are vulnerable to denial of service attacks, and 99.84% of MQTT- and XMPP-speaking backends use insecure transport protocols (only 0.16% adopt TLS, of which 70.93% adopt a vulnerable version).

## 1 Introduction

The number of active Internet of Things (IoT) devices is forecasted to reach more than 29 billion by 2030 [105]. They assist people with health sensors and improve living conditions in smart homes (e.g., alarms and thermostats) and smart cities (e.g., air quality monitors). These devices typically rely on backends, that is, *servers, commonly deployed in the cloud, that store and process data as well as control the connected devices*. Backends thus play a vital role in the IoT ecosystem, and their security is crucial. Vulnerable backends enable a variety of attacks, such as information exfiltration or denial of service (DoS). Such attacks are far from hypothetical and unprecedented in their scale due to IoT devices' pervasiveness [33, 95]. Sabetan [85] discovered that over 40,000 Nexx's Smart Garage doors were vulnerable due to a misconfigured backend. An attacker could have opened any garage door from anywhere in the world as a single insecure password was used to protect data for all customers. Several issues also arose with connected kids' devices and their insecure backends [36, 37, 55]. CloudPets allowed parents and kids to record and receive audio files through Internet-connected plush toys. However, over 2M recordings of over 800k users containing personal conversations were publicly accessible as they were stored in a MongoDB database with hardly any authentication [38, 58].

There are only few standardized attempts for securing the IoT ecosystem, such as Manufacturer Usage Descriptions (MUDs) [26, 28, 50], but they are only high-level descriptions and not yet deployed [42, 59]. Incorrect and poor documentation also leads to vulnerabilities. Jia et al. [43] showed how 26 out of the 38 "best practices" in the official developer guide of Amazon Web Services (AWS), the leading IoT cloud platform, introduced vulnerabilities.

Albeit security is a primary concern for developers [27, 100], the vast amount of communication protocols and the heterogeneity of IoT environments make it difficult for them to fully comprehend the overall situation. The many protocol standards and the plethora of IoT devices, from pacemakers to smart refrigerators, further complicate the situation. Each device and deployment has different requirements and resource constraints, making developing and enforcing security and privacy measures challenging.

Previous studies focused on identifying device-based vulnerabilities or proposed protocol-specific solutions. In addition to extensive studies on devices' susceptibility to malware [1, 3, 22], some approaches leveraged companion apps to improve the scalability of analyses of the IoT ecosystem, but, again, mainly focused on device security [15, 82, 94]. There is a need for comprehensive studies assessing the security of IoT backends as they represent an easy entry point for attackers and allow the escalation of attacks to any devices connecting to them [111]. Backends are the backbone of the IoT ecosystem and ensuring their security is critical [29].

In this paper, we fill this gap by measuring the security posture of publicly accessible IoT backends, that is, servers speaking IoT-focused protocols, at scale. We investigate three application-layer messaging protocols widely adopted in the IoT: Message Queue Telemetry Transport (MQTT), Constrained Application Protocol (CoAP), and Extensible Messaging and Presence Protocol (XMPP).

Maggi et al. [54] and Palmieri et al. [76] provided first insights into the security of MQTT and CoAP backends, identifying exposure of sensitive information, for example, ambulances leaking their geographical locations. Our work complements and substantially expands on their work by scrutinizing more and different classes of vulnerabilities, such as DoS, and characterizing the security posture of real-world IoT backend deployments at scale.

Note that we do not investigate HTTP backends because determining if it is used in the context of the IoT is extremely challenging as it requires a semantic understanding of the exposed API, demanding analyses that do not scale and require invasive and ethically questionable probing. These analyses would have to be performed over large parts of the Internet due to the popularity of HTTP when only a small to negligible number of HTTP backends are involved in the IoT. Generally, HTTP has shifted from an application-layer protocol for the web to a common transport protocol for many applications (e.g., DNS-over-HTTPS [11, 35], HbbTV [106, 107]). Thus, we focus on protocols that devices themselves “speak” and that are tailored for the IoT.

We leverage Shodan to crawl public backends since existing datasets contain insufficient IoT backends, preventing us from painting a complete picture of the ecosystem. We can only gather 223 backends from prior datasets speaking the selected IoT protocols, while we collect and analyze over 337,000 backends. We infer the deployed software versions, list their exposed topics and resources, and test if security and privacy measures are in place, e.g., authentication or Transport Layer Security (TLS). We discover thousands of vulnerable backends and cases of sensitive data exposure. Furthermore, we repeat our evaluation over time to provide a longitudinal view of the security and privacy of IoT backends.

Overall, we make the following contributions:

- We gather a dataset of 337,464 backends speaking MQTT, CoAP, or XMPP, and record 10.6GB of network traffic.
- We evaluate backends’ security and privacy posture at scale, studying misconfigurations and vulnerabilities. We discover critical security issues, like weak authentication and the potential for amplification attacks that can enable DDoS.
- We investigate TLS adoption, analyzing their use of insecure versions and cryptography, and expired certificates.
- We report our findings through a Coordinated Vulnerability Disclosure (CVD) process to the backends operators and provide guidance to support their remediation efforts.
- We repeat our analysis and show that, despite improvements and our disclosure, some backends exhibit worse security and are affected by more vulnerabilities over time.

**Ethical Considerations and Disclosure.** Naturally, our measurement prompts ethics questions. We describe our precautions to prevent potential harms in Appendix A. Our institution’s ERB has reviewed and approved our study. We discuss our CVD process with the Dutch CERT and Cyber Security Center in Appendix B.

**Artifacts.** Due to the sensitive nature of the data, we make our datasets available to other researchers on request. Our source code is available at https://github.com/SecPriv/IoTBackends.

## 2 Background

IoT messaging protocols offer tailored functionalities that account for the characteristics of the devices, their resources, and network constraints. This makes them particularly interesting from a security perspective, as these trade-offs impact confidentiality, integrity, and availability. Following, we introduce the most widely adopted IoT protocols. We exclude HTTP from our study because discerning whether an HTTP backend serves (only) IoT content does not scale and requires invasive and ethically questionable analyses, while HTTP is also mostly spoken by companion apps and mostly used as a transport protocol for other protocols on top of it [94].

**Message Queue Telemetry Transport (MQTT).** MQTT is a lightweight publish/subscribe IoT messaging protocol standardized by OASIS [9, 10]. Its operation revolves around three entities: the broker, that is, the *backend*, and one or more publisher and subscriber clients, for example, IoT devices. The broker is a centralized entity that receives PUBLISH messages. It routes them based on subscriptions, access control rules, and a Quality of Service (QoS) that can be both associated with the sender or the receivers (i.e., 0 = at most once, 1 = at least once, 2 = exactly once). Publishers typically open a connection with the broker via a CONNECT packet and send one or more messages on specific topics by indicating a QoS. Topics are hierarchically organized paths and can include wildcards (“+” and “#” respectively). Subscribers subscribe to one or more topics via SUBSCRIBE packets.

**Constrained Application Protocol (CoAP).** CoAP is a REST-oriented IoT messaging protocol focusing on resource-constrained Machine-2-Machine (M2M) communication [12]. A client requests a resource from a server, that is, a *backend*, via its URI (Uniform Resource Identifier). The request can be a confirmable (CON) or non-confirmable message (NON). The protocol is asynchronous and relies on UDP (User Datagram Protocol) as the transport, which is connection-less and does not provide a re-transmission mechanism. CON packets guarantee a certain reliability. CoAP methods mirror HTTP methods: GET to fetch a resource, POST, PUT, and DELETE to create, update and delete it. CoAP can be used with Datagram TLS (DTLS) to improve security, known as *coaps*, and it supports four security modes: No Security, Pre-Shared Key, Raw Public Key, and Certificates. Their choice is dictated by resource availability, security requirements, and deployment (e.g., Internet access).

**Extensible Messaging and Presence Protocol (XMPP).** XMPP is an application profile of the Extensible Markup Language (XML). It enables the near-real-time exchange of structured and extensible data between two or more network entities [87, 89], building on a distributed client/server model to exchange “stanzas” (XML data). XMPP relies on TCP and open XML streams to exchange stanzas. It allows client-to-server and server-to-server communication. In the latter, one server acts as a client with the difference that its addresses are known a priori. We consider XMPP servers as *backends*.

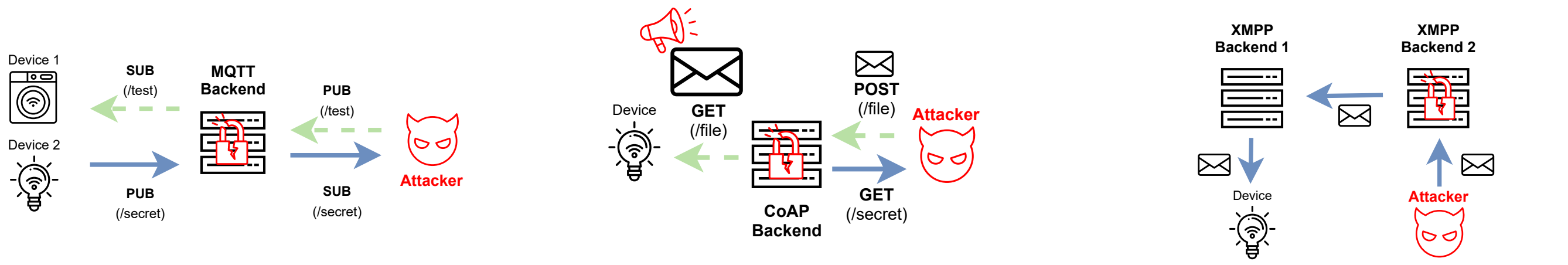

(a) MQTT (b) CoAP (c) XMPP

**Figure 1: Attack Vectors for the Studied IoT Protocols.** We consider three different architectures: Publish-Subscribe (MQTT), Client-Backend (CoAP), and Distributed Client-Backend (XMPP). The colored arrows signify different flows.

## 2.1 Security Concerns with IoT Protocols

*2.1.1 Threats per Protocol.* We focus on distinct threats for MQTT, CoAP, and XMPP, shown in Figure 1. We consider other attacks, like breaking cryptographic ciphers, out of scope.

**MQTT.** Authentication represents a problematic issue for MQTT. Enforcing authentication via credentials or certificates is possible but not required. Authorization-wise, only simple Access Control Lists (ACLs) are supported. Data integrity can be enforced via TLS. Authentication is not used by default, even when brokers support password-based authentication. Attackers can sniff credentials by intercepting CONNECT packets as they are sent in plaintext (no TLS). Successfully connecting to a broker puts confidentiality at risk. Attackers may also be able to subscribe to all topics with wildcards and potentially access sensitive information.

Data integrity is at risk if attackers can intercept messages, replay them, or alter their payloads. They could perform confused deputy attacks by modifying firmware update message files or references.

**CoAP.** Due to UDP's nature, CoAP is vulnerable to IP address spoofing attacks. UDP cannot authenticate a communication partner; endpoints cannot verify if a packet truly originated from the claimed source IP address. An attacker can send a request with a spoofed IP address, and the backend, since it might trust the source address, might process the request and respond to the spoofed IP address. The server then acts as a reflector [54]. CoAP is susceptible to IP spoofing only when not adopting adequate authentication. For example, DTLS enables verification of communication parties.

CoAP can also enable amplification attacks. As it uses the request/response model, the server responds when receiving a request. However, the response size can be substantially larger than the request. Adversaries can send many small request packets to generate large response packets. In turn, amplification combined with IP address spoofing allows attackers to launch DoS attacks at victims.

**XMPP.** XMPP uses TLS with the STARTTLS extension for session encryption, protecting communication from eavesdropping and tampering and allowing to upgrade existing insecure connections to secure TLS ones. However, a downside of STARTTLS is that it makes XMPP vulnerable to downgrade attacks, enabling Monster-in-the-Middle (MITM) attackers to read and modify XML stanzas [65].

Other issues lie in how encryption is used. A stanza can be sent over multiple XML streams, but there is no guarantee that all streams are encrypted. Therefore, end-to-end encryption is vital to protect the stanza on every hop, but the XMPP community does not yet provide a suitable technology [88].

# 3 Threat Model

Attackers have the following capabilities: They can connect to the vulnerable backends, which is possible in our case because all backends are publicly reachable on the Internet. Without a valid connection, the attacker would be unable to connect to them and exploit the vulnerabilities. They have the required knowledge about the protocols and vulnerabilities to successfully exploit them. For example, some weaknesses require specific message formats or payloads, such as, allowing "#" for listing all MQTT topics.

**Information Leakage.** IoT backends can unveil different types of information, ranging from software information (e.g., library versions) to exchanged messages. In the worst case, an unsecured MQTT broker can expose health monitors' (e.g., insulin pumps) data and patients' Personally Identifiable Information (PII) [54]. Attackers can also leverage other types of information to gain further access: They can exploit known vulnerable software versions (e.g., leading to crashes or taking over control) or target individuals. This threat class is connected to *Weak Authentication*, as an attacker who bypasses authentication can often access unauthorized data.

**Weak Authentication.** Weak authentication mechanisms are a known problem of the IoT. The resource-constrained nature of IoT devices makes adopting security features costly, and often, developers rely on security-by-obscurity, assuming the non-triviality of reversing devices' firmware [30]. Even when security best practices are adopted, such as TLS, they are often incorrectly implemented. Paracha et al. showed how most devices they tested use old or insecure protocol versions and cipher suites, and lack certificate validation [77]. Bypassing authentication can also allow attackers to gain full control of a system, allowing them to gain access to (sensitive) data, send crafted messages to clients, or spam fake data.

**Denial of Service (DoS).** The problem of DoS attacks is two-fold: (1) an attacker can target an IoT backend, or (2) the IoT devices. In the former case, a malicious actor can impede communication between clients by taking down the backend. In the latter case, the clients would become unresponsive or crash; thus, it would not be able to perform its task. Considering the potentially critical settings of some IoT deployments, like power plants, such attacks could lead to power blackouts in a geographical area. Moreover, backends can act as amplifying reflectors for distributed DoS (DDoS) attacks when their response is larger than the request size. Given the limited resources of IoT devices, even a moderate amplification factor can overwhelm devices. If the victim device is medical (e.g., an insulin pump), then a DoS can cause its users serious life-threatening harm.

## 4 Motivation

We discuss two motivating examples that highlight the importance of our study to assess the (in)security of IoT backends. We show how weak authentication and information leakage can pose serious threats to users' security and privacy, as introduced in Section 3.

**Methodology.** Existing research used static analysis for IoT devices' companion apps to spot vulnerabilities in the devices and to extract their backends [43, 44, 45, 60, 62, 94]. We use IotFlow [94] to statically reconstruct the backends that IoT companion apps contact and investigate two example apps with insecure backends based on the associated devices and the data they exchange.

**Heart & Lung Monitor.** The first example is an MQTT broker for a wearable smart device that monitors a user's lungs and heart. We connect to the broker and subscribe to the wildcard topics "#" and "$SYS/*" with QoS 0 to avoid acknowledging messages intended for other connected clients and listen only passively. We retain only topic names and "$SYS/*" payloads and do not record any other information. We find that messages reveal PII of users, such as name, age, and gender, in addition to several health indicators, such as heart and breathing rate, and precise geographical location. An attacker listening to incoming messages can not only precisely identify users and their geolocation, but also alter health indicators. Furthermore, from "$SYS/*" payloads, we can infer the mosquitto library, version 1.4.15, released in February 2018. This version suffers from authorization and DoS vulnerabilities [68, 69, 70].

**Smart Car Dongle.** We reconstructed the second backend from an app associated with a dongle to bring smart car features to regular cars. Its functionalities include real-time geolocation monitoring, engine monitoring for anomalies, and anti-theft alarms. When connecting to the broker, we discovered that messages reveal sensitive information about the cars' brand and type, location, fuel consumption statistics, and speed. The broker also unveils users' email addresses and the anti-theft alarm's status (i.e., On/Off). The anti-theft status combined with cars' type and precise geolocation makes a perfect list of valuable cars for thieves to target.

**Summary.** We show anonymized example messages for the two apps in Appendix C. In both cases, weak authentication, or rather a lack of authentication, allows arbitrary anonymous users to connect and read messages. The exposed data we identified is clearly sensitive, highlighting the severity of the problem and potential consequences. We first responsibly disclosed the issues to the developers in May 2023. The developers of the health monitor app replied to our second email. They have deprecated the identified backend and are moving their services to an AWS-managed backend. The legacy backend remains available for backward compatibility until the remaining users update the app. Unfortunately, we have not received any response from the car dongle manufacturer, even after repeated follow-up emails.

## 5 IoT Backend Datasets

The two motivating examples already show how misconfigured and vulnerable backends can impact users of IoT devices. There is a clear need for a comprehensive analysis of publicly accessible IoT backends at scale to identify and understand ecosystem problems and propose informed and viable solutions.

**Table 1: Datasets Used in Our Study.** We report the number of unique IoT devices used to capture the network traffic, the number of unique backends (based on IP or domain), and the number of IoT-specific backends (MQTT, CoAP, XMPP). We also collected a dataset from Shodan, thus, only the IoT backends are available.

| Dataset | Where? | When? | # IoT Devices | Unique Backends All | Unique Backends # IoT (%) |
|---|---|---|---|---|---|
| IoT Sentinel [57] | FI | 2016 | 31 | 101 | 0 (0.00%) |
| UNSW [99] | AU | 2016 | 28 | 9,610 | 125 (1.30%) |
| IoTLS [77] | US | 2018–2020 | 40 | 1,495 | 68 (4.55%) |
| YourThings [4] | US | Q1 2018 | 45 | 7,172 | 32 (0.45%) |
| Mon(IoT)r [84] | US+UK | Q1 2019 | 81 | 3,570 | 17 (0.48%) |
| IoTFinder [80] | US | 09/2019 | 53 | 7 | 0 (0.00%) |
| PingPong [108] | US | 11/2019 | 19 | 6,848 | 25 (0.36%) |
| HomeSnitch [74] | US | Q1 2020 | 24 | 1,436 | 57 (3.97%) |
| Edge IIoT [31] | DZ | 01/2022 | >10 | 38 | 0 (0.00%) |
| SHODAN-22 | - | 07/2022 | - | - | 901,295 (-) |

To collect a comprehensive dataset of real-world backends that speak IoT protocols, we investigate datasets from prior work and collect our own. Table 1 provides an overview of our datasets.

**Existing IoT Traffic Datasets.** We collect nine IoT traffic datasets from prior work. We extract IPs, associated ports, and DNS information from traffic dumps (pcaps) and map IPs to domains. IP addresses can vary over time, but this is less likely for domain names, which means they can yield more accurate results. When we cannot match IP and domain, we retrieve the (historic) reverse DNS names for the IP via Shodan [98]. We acknowledge that Shodan's databases may be incomplete or outdated, which is a known limitation of it and related approaches [112]. For our analysis, we consider unique backends for which the associated port is a default IoT protocol port, namely 1883 and 8883 for MQTT, 5683 and 5684 for CoAP(s), and 5222, 5269, 5280, and 5298 for XMPP. Overall, the datasets contain only 45 MQTT, 3 CoAP, and 175 XMPP unique backends. The datasets contain little network traffic for the three IoT protocols, highlighting their partial and limited coverage of the IoT ecosystem.

*Note*: We tried to include the dataset by Saidi et al. [86], but they could not share it because it contains proprietary data from Farsight. This makes it impossible for us to reproduce, validate, or extend their work.

**Shodan Crawl (SHODAN-22).** Given the limited coverage of the IoT backend ecosystem of previous datasets, which do not allow a large-scale analysis of backends, we crawl Shodan [98] for Internet-connected devices. We search Shodan in the last week of July 2022 for the keywords *mqtt, coap(s), and xmpp*, and find 425,571 MQTT, 474,878 CoAP, 4 CoAPs, and 702 XMPP results, without restricting us to standard ports. We store IPs, available hostnames, and ports and add extra information, such as connection codes.

### 5.1 Dataset Augmentation

We augmented our dataset with additional information, namely the geographical location and whether backends are hosted on widely adopted cloud platforms. We use Shodan and `WHOIS` to determine the country of the backend. We analyze backends at a country-level granularity, and thus, our geolocalization is sufficiently accurate, as country misclassifications are rare. We also gathered the IP address ranges for ten major cloud providers that offer managed IoT services: Amazon Web Services (AWS), Google, Cloudflare, Microsoft Azure, Alibaba, Oracle, IBM, DigitalOcean, Yandex, and Salesforce [2, 5, 17, 24, 32, 40, 56, 75, 92, 113]. We determined if each backend's IP address belonged to one of the providers. If it does not belong to one, we classify its provider as *Other*. IP address ranges of cloud providers might change over time, which is why we include them in our artifact. Finally, we use the regular expression by Saidi et al. [86] to distinguish between self-hosted and managed AWS backends (i.e., hosted on the AWS cloud vs. managed by AWS).

**Countries.** Most SHODAN-22 IoT backends are located in South Korea, followed by China and the Philippines. Interestingly, breaking this down at the protocol level, most MQTT backends are located in South Korea (276,100, 64.88%), followed by China (46,391, 10.90%) and Japan (18,204, 4.28%). For CoAP, most backends are in the Philippines (167,849, 35.35%), followed by Russia (104,639, 22.03%) and China (80,619, 16.98%). We explain geolocation trends based on where most manufacturers and vendors are located. Shadowserver observed a similar trend for CoAP backends [47, 96]: While they only probe for services exposing the resource *.well-known/core*, they saw the same distribution, with the Philippines, China, and Russia making up nearly 93%. Maggi et al.'s study [54] showed a different country distribution for MQTT and CoAP backends in 2018. They identified numerous backends in the US. One reason could be dynamic changes in IoT backends' locations over time, aligning with major vendors' (re)locations, or increased white-labeling of IoT devices. Another reason could be that some backends are not IoT but have adopted IoT protocols (e.g., due to similar resource constraints). Given ethical considerations, a precise distinction is impossible, as it would require invasive measurements, making our data an upper bound (see Section 8). Moreover, other backends using IoT protocols might suffer from the same issues as IoT backends, which is also interesting to study.

**Providers.** We identified the hosting providers for 31,785 backends, of which 25,146 (79.00%) belong to AWS, 3,381 (10.62%) to Azure, 2,292 (7.20%) to Google, 674 (2.12%) to Oracle, 282 (0.89%) to Alibaba and the remaining ten 10 (0.03%) to Cloudflare. Our trend reflects the market share ranking for cloud providers in 2022, with AWS, Azure, and Google being the top three providers [46]. Particularly interesting is that AWS hosted almost 80% of backends with 34% market share, while Azure and Google account for 11% and 7% backends at 21% and 11% market share, respectively. We found no backends hosted on DigitalOcean, Yandex, IBM, or Salesforce.

## 6 Large-Scale Security Assessment

Following, we describe our methodology for our large-scale security measurement and assessment of the identified IoT backends. We discuss the vulnerabilities and weaknesses that we study and explain their relevance. We also expand on how we implement our approach for each messaging protocol. The protocols that we investigate have different architectures and face different threats (discussed in Section 2 and illustrated in Figure 1): Publish-Subscribe (MQTT), Client-Server (CoAP), and Distributed Client-Server (XMPP). Therefore, they require different measurement approaches, and we describe our methodology and results per protocol. We also investigate TLS usage for the TCP-based protocols MQTT and XMPP. TLS is a widely adopted security measure for TCP-based protocols. Since most CoAP deployments rely on UDP, TLS cannot be used and DTLS is required. However, DTLS analysis remains an open research area, and only 4 of 474,882 CoAP backends even support DTLS (0.0008%), thus, we leave it for future work.

Considering the low number of IoT backends from existing datasets and that most were unreachable as of September 2022 (93% were unreachable; we could only connect to one MQTT, zero CoAP, and 14 XMPP backends), we exclude them from our analysis for clarity and readability. We attempted to include additional backends by extracting them from companion apps utilizing IotFlow [94], but we could only identify less than 100 backends in over 4,000 manually verified companion apps. Likely, most IoT devices do not rely on the mobile app but communicate with IoT backends directly, which is intuitive, as otherwise, they could stop functioning when the phone is unavailable. Thus, we perform our security analyses on the SHODAN-22 backends.

For our analyses, we act as unauthenticated users without prior knowledge of the target backend. Our measurements do not require privileges (e.g., admin) or authentication. While we can identify vulnerabilities without requiring access privileges, this does not imply that they are always exploitable. We also cannot test for actual exploitability as this would clearly violate ethics and possibly disrupt or compromise services.

Finally, we repeat our analyses after one year to determine if operators improved the security of vulnerable IoT backends over time. We also evaluate the stability of our dataset because IoT traffic datasets tend to age quickly, as discussed in Section 6.7.

### 6.1 General Approach

We identified the main security issues in IoT backends by leveraging prior work and following our classification in Section 3. Prior work identified insufficient authentication and authorization measures as key issues [29, 61], as well as the adoption of outdated libraries with known vulnerabilities [54], helping us understand how weaknesses in our threat classes can be discovered in the backends. We then listed all potential attacks for each class, such as using default credentials in case of *Weak Authentication*. Particular caution is required because we interact with real-world backends, which might provide critical functionalities. For this reason, after thoroughly studying and evaluating the possible ethical implications of our study (see Appendix A), we removed all attacks from our list for which existence testing *could* alter the backends' correct functioning. For example, we decided against testing for known/default credentials as (1) it could cause DoS in case requests are too frequent or possibly prevent authentication by a legitimate user because of too many unsuccessful authentication attempts, and (2) it likely would also cross legal boundaries in some jurisdiction, as we would access a system that could be considered legally protected.[1]

Importantly, we test for specifically selected vulnerabilities and threats, meaning our results may not generalize to other vulnerabilities or threats. Nevertheless, our results provide new insight into the security state for IoT backends, as the classes of vulnerabilities we analyze are general and well-known threats. Moreover, our analysis allows us to investigate how the backends' operators are considering and approaching security because the specific vulnerabilities and threats we analyze differ in how long mitigations and patches have been available or known, how easy they are to deploy, and in their severity/impact.

[1]We initially considered default credentials, and an expert assessment indicated that this would be legal in at least some of our jurisdictions, as protection with default credentials should not be considered adequate protection. However, our research touches many jurisdictions, and the scientific benefits of doing so are limited.

We first collect general information about the backends and identify the libraries they use, including their versions. This information can tell us whether a system under analysis adopts the most recent security patches. In this way, we can determine if backends suffer from known vulnerabilities. When available, we also collect complementary information, like the authentication mechanisms and the number of connected clients. We then investigate possible information leakage from these backends. First, we analyze the communication structures and determine how messages and resources can be retrieved. We then scrutinize data for privacy issues and confidentiality violations. Considering that we act as unauthenticated users, we should not have access to any sensitive information. Figure 2 illustrates our complete methodology.

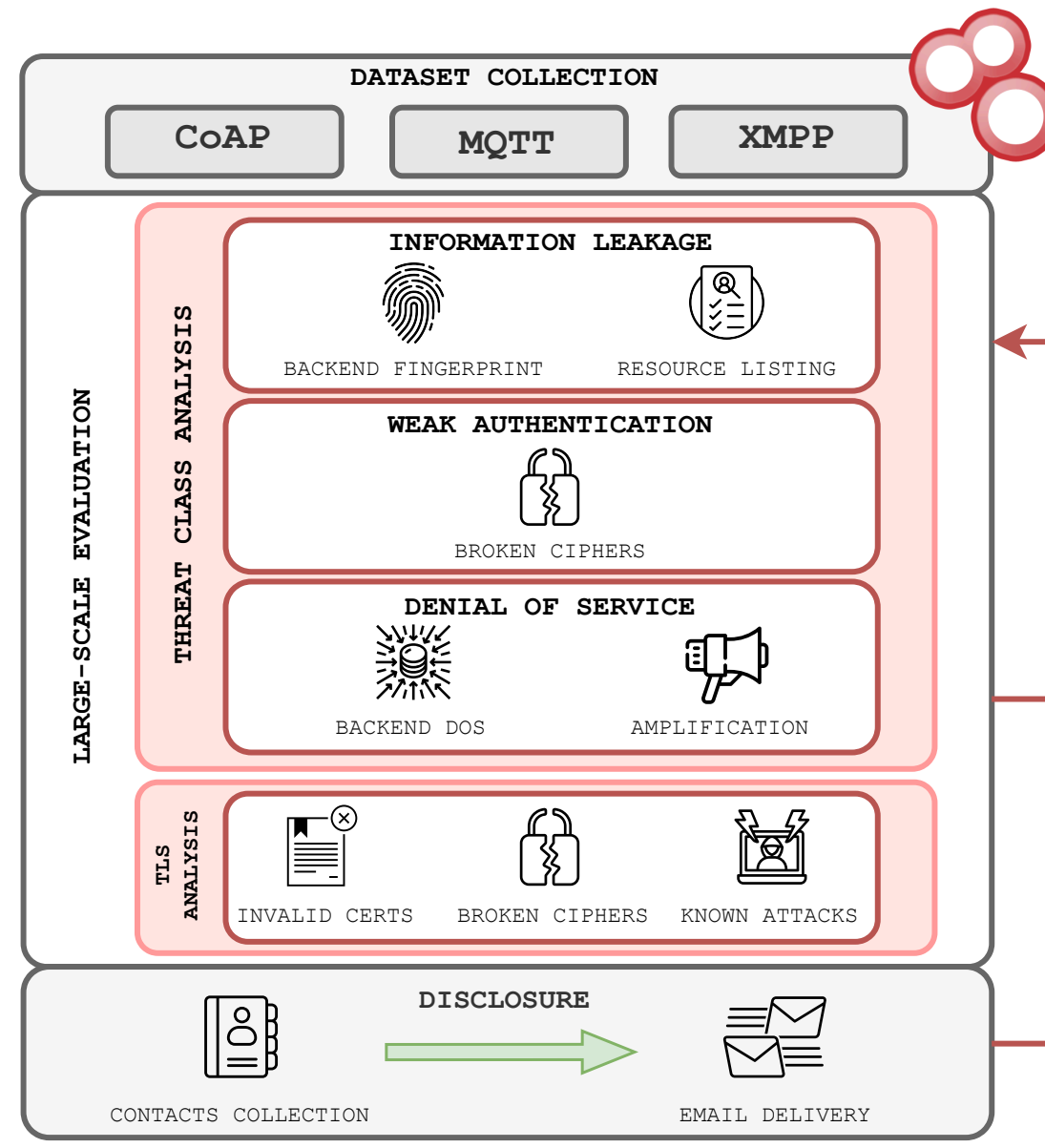

**Figure 2: Methodology Overview.** We first collect our backends' dataset via Shodan. Then, we perform our large-scale evaluation for the identified threat classes. For TCP-based protocols (i.e., MQTT and XMPP), we also evaluate TLS adoption and implementation. We repeat our analysis over time to understand how security posture evolves, and to consider the impact of our coordinated disclosure.

## 6.2 Measurement Setup

We perform our initial analyses on 425,571 MQTT, 474,882 CoAP, and 702 XMPP backends between August and November 2022 from the Netherlands. We perform all tests on a Ubuntu 22.04 virtual machine (VM) with eight CPU cores (Intel Xeon Silver 4110) and 32 GiB memory, with a timeout of 60 seconds per backend. We analyze up to 10 MQTT and CoAP backends in parallel to minimize network input/output wait times. To foster further analysis, we record all our analysis traffic (10.6 GiB) using `tshark`, filtered on the backends' IP, which we make available upon request.

**Reproducibility and Tooling.** For some analyses, we adopt and adapt existing security tools. This is motivated by their stability and reliability, making it easier to reproduce our results. We recommend IoT developers adopt them for vulnerability testing of their deployments, making our results more approachable and understandable while fostering scalability. We report the tools and how we utilized them in the approach paragraph for each protocol.

## 6.3 MQTT

**Approach.** We identify MQTT backends' software version and analyze existing topic names using the *Paho* library (v1.6.1) [51]. We complement the MQTT backends we gathered in SHODAN-22 with connection codes: If the connection is successfully established, it returns code 0. Otherwise, codes 1 to 5 mean an error related to connection parameters. We define three cases based on the return code: for code 0, we can connect and proceed; for code 1, we try a different protocol version (e.g., MQTTv5); for codes 2 to 4, we mark it as available but requiring authentication or authorization. We connect to them via IP/hostname, port, and protocol version.

After a successful connection, we subscribe to the wildcards "$SYS/#" and "#" with QoS 0 to avoid acknowledging messages intended for other clients. With QoS 0, we do not acknowledge messages, and the broker will re-attempt sending them to other (legitimate) clients until they are acknowledged (by them). We listen for incoming messages for 40 seconds. We record the topic names and count the received messages. We do not record message payloads, except for "$SYS/" topics, which provide relevant information for backend fingerprinting and security analyses, like the broker version or the number of connected clients. While some topic names could include potentially sensitive information, we only record minimal metadata and deem this risk acceptable.

We test for two vulnerabilities via *cotopaxi* (v1.6.0) [93], an IoT pentesting tool by Samsung: CVE-2019-9749 [72] causing the crash of Fluent Bit brokers (version < 1.0.6) and CVE-2018-19417 [67], a stack-based buffer overflow in Contiki OS MQTT brokers (version ≤ 4.1), by checking the broker version. The former vulnerability makes the broker unavailable via a DoS attack, thus impeding communication between IoT devices. The latter allows remote code execution on the broker and full memory access.

**Results.** Overall, we successfully connected to 251,382 out of 425,571 (59.07%) MQTT backends, which we consider the baseline.

### 6.3.1 *Weak Authentication.*

We observed 12,071 backends (4.80%) that returned a code 4 upon connection, which indicates that they use username-password-based authentication. These credentials are sent in plaintext, enabling eavesdroppers to intercept them and connect to the MQTT broker if TLS is not used properly.

### 6.3.2 *Information Leakage.*

**Demographics.** MQTT allows us to identify the number of unique clients, that is, IoT devices, connected to a broker. Brokers with more clients could unveil more information as more messages might be exchanged, and they can be more impactful targets for DoS attacks. The average number of connected clients is 11.7 ($\sigma$=538.04), peaking at 33,134 for a single broker. Per geographical region and provider, the average number of connected clients is less than ten, but with long and dense outlier tails (see Figures 5 and 6 in Appendix D).

**Software.** We identify, when possible, the library and its version adopted by the MQTT broker via messages sent on the topic "$SYS/broker/version." If this topic is restricted or we cannot connect to the backend, we cannot infer the software version. We analyzed version information for 22,986 backends. After removing artifacts, almost all brokers (22,978, 9.14%) use mosquitto, an open-source MQTT broker by Eclipse [52]. Worryingly, 10,627 backends (4.23%) adopt a library version that was released more than five years ago (version < 1.5). Outdated library versions can indicate poor security practices and allow the exploitation of vulnerabilities addressed in later versions. Updating versions is, however, not always possible, as some devices might not allow updates. Investigating the vulnerabilities that old mosquitto versions suffer from, we find that versions 1.0 to 1.4.15 (10,627 backends, 4.23%) are vulnerable to a null pointer dereference that can cause a broker crash (DoS) [68]. 11,804 backends (4.70%) use versions older than 1.5.5, which exposes them to two vulnerabilities: Malformed data contained in a password file [70] or an empty ACL file [69] allows attackers to circumvent authentication and authorization checks and access all the information exchanged through the broker.

**Topics.** We perform an in-depth analysis of the topic names we observed to understand what type of data is exchanged. For example, to determine whether it is IoT device data or unrelated. We observed 1,766,804 unique topics (average 39.45 per backend, peak of 36,254 unique topics for one backend) of which 50,167 are "$SYS/" topics. We use the zero-shot classification model by Laurer et al. [49] to analyze if the remaining 1,717,765 topic names fall into nine major IoT-related categories: *health*, *home*, *security*, *update*, *sensor*, *location*, *industry*, *transportation* and *identifier*. The classifier assigns a score for each topic name and category from 0 (no match) to 1 (perfect match). We require a minimum score of 0.85, as recommended by Laurer et al., for a topic to match a category. Topics may match zero categories (no score $\geq$ 0.85) or multiple (2 or more scores $\geq$ 0.85). If a topic matches multiple categories, we count it only for the top-scoring one. Overall, we classify 697,818 topics (40.62%) into the nine IoT-related categories (see Figure 3), providing strong evidence that the brokers are indeed used for IoT communication.

Manually analyzing the categorized results, we identified topics that might contain sensitive information that should not be publicly accessible. Following, we redact some of the information to avoid identifiability. For example, *cmnd/-/mqttpassword* appears to unveil the MQTT password, while *Security/GarageDoorFront* and *-/Living_-Room/Front_Door_Sensor/-/Home_Security/Cover_status* appear to allow controlling smart home devices. Several other topics are potentially associated with a firmware update functionality. Delivering an update over plaintext channels is a major security concern for smart devices. If the update is not cryptographically signed, then an attacker can replace it with custom malicious firmware [41].

**Case Studies.** We investigate two specific backends in more depth: (1) the backend for which we collected the most topics and (2) the broker with the most connected clients.

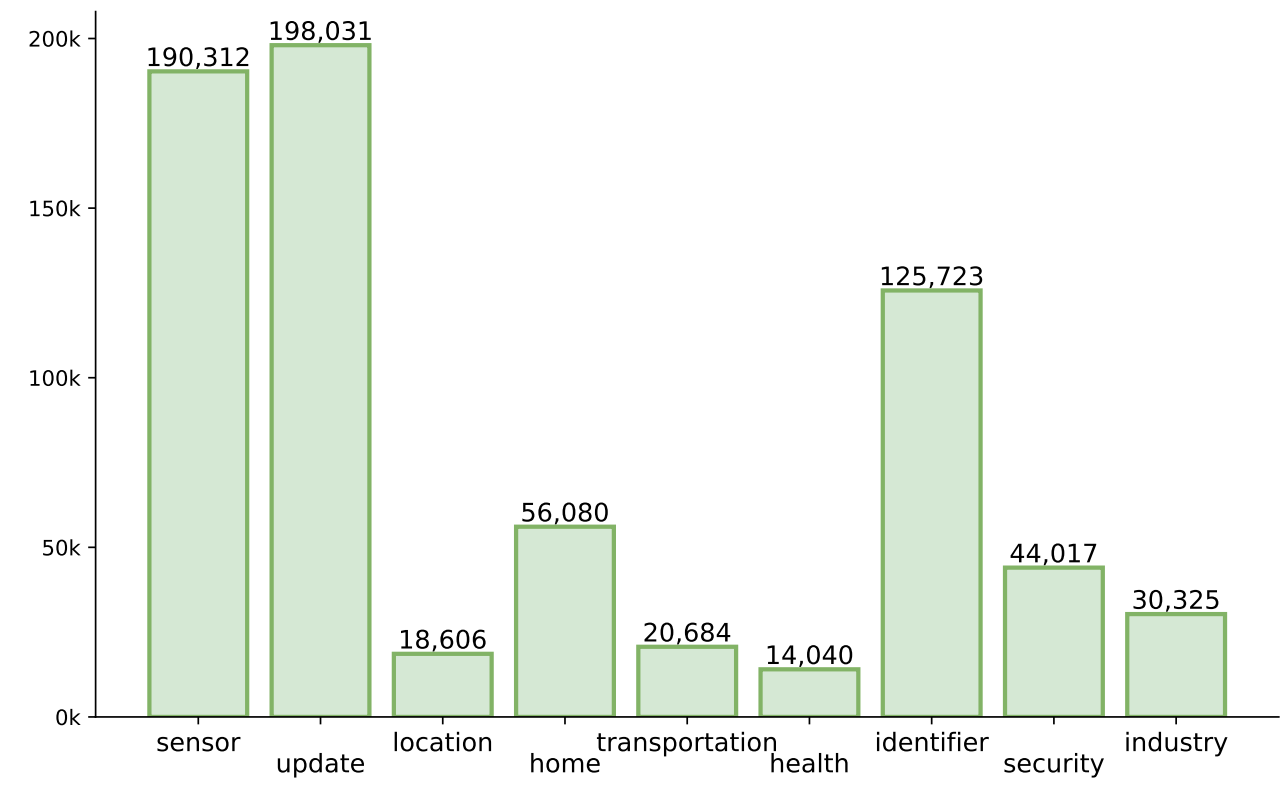

**Figure 3: Classification Distribution of 697,818 MQTT Topics.** The most common topic categories are *sensor* and *update*, with *security*-associated topics also occurring frequently.

Analyzing the former's topics, it appears to correspond to a power plant in China: Two of its topics are *signal-values/admin/-/_station_efficient/power-facility-value* and *-/station_management/alarmSeverityMap*. Exposing such sensitive information is a clear and severe security threat that could have disastrous consequences, considering how critical the power grid to modern society is.

For the latter, we could not collect any topics. However, it relies on an outdated version of the mosquitto (v1.4.14 from July 2017), which suffers from two authentication and authorization vulnerabilities. Therefore, attackers might be able to exchange messages with many IoT devices (33,134) by exploiting the vulnerabilities and circumventing authorization. This exposes the devices to multiple risks. For example, an attacker could send carefully crafted messages to devices and exploit other vulnerabilities to make them join a botnet, or eavesdrop on their communication.

*6.3.3 Denial of Service.* Finally, we investigate if backends suffer from two known vulnerabilities that can be used to launch DoS attacks, namely CVE-2019-9749 and CVE-2018-19417. We identified 214 and 196 backends affected by CVE-2019-9749 and CVE-2018-19417, respectively. CVE-2018-19417 allows remote code execution, including letting attackers take the broker offline or disconnect clients, while CVE-2019-9749 crashes the broker directly.

**Recommendations:** We encourage developers to update their broker libraries, thus fixing known vulnerabilities in older versions. We found vulnerabilities affecting broker versions older than five years, signaling bad security and update practices. In some cases, updating might not be trivial, for example, when IoT devices are running old and broken software versions that are incompatible with backend updates.

We suggest that stakeholders carefully evaluate what information needs to be accessible and by whom and to adopt authentication measures and ACLs. With mosquitto, the most widely used MQTT broker we found, this can be done with a text file that lists each user's permission for specific topics.

Encrypted communication (e.g., TLS) should be used when possible. Finally, if the broker does not have to be accessible to the entire Internet, access should be restricted via a firewall.

## 6.4 CoAP

**Approach.** We identify the adopted software and the number of connected clients of CoAP backends with *cotopaxi* (v1.6.0) [93]. We compile a list of 30 resources we check for: *helloWorld, test, login, admin, administrator, adm, .passwd, passwd, history, certificates, logout, password, log, logs, about, actions, advanced, auth, backup, .well-known/core, .history, certs, config, configuration, data, dev, files, help, resources,* and *items*. For example, a CoAP backend exposing the `password` resource without protection measures might indicate a sensitive information leak. We also include `.well-known/core`, a default URI used as an entry-point for listing the resources hosted by a backend (but not always available). We perform a HEAD request for each resource with a sleep of one second and look at the return code. We mark the resource as available if we receive the return codes 2.05 (Content) or 2.03 (Valid).

We test for two traffic amplification vulnerabilities with *cotopaxi*, CVE-2019-9750 [73] targeting IoTivity (an open-source framework for device-to-device connectivity) and ZYXEL_000 affecting Zyxel Keenetic routers. For both, we send a message and check the response size. If the amplification factor (AF) is greater than 100%, we flag the backend as vulnerable. Such vulnerabilities allow attackers to abuse CoAP backends as reflectors to take down connected IoT devices. We also test for the DoS vulnerability CVE-2018-12679 [71]. In this case, the target of the DoS attack is the backend itself, which can no longer serve content to its clients.

**Results.** We successfully connected to 85,957 out of 474,882 (18.10%) CoAP backends, which we consider the baseline.

#### 6.4.1 *Information Leakage.*

**Software.** We identify the software for 2,864 backends: 1,886 (2.19%) adopt *coap-rs* [19], followed by 932 (1.08%) *FreeCoAP* [20], and 19 *aiocoap* [6] instances.

**Exposed Resources.** All 30 resources we defined are available. The resource available across most backends is `.well-known/core` (767 backends, 0.89%), followed by `test` (10 backends) and `help` (6 backends). The resource `admin` is publicly accessible for six backends and `password` for seven. Exposing potentially sensitive content without further security measures is a severe security threat. In four cases, the resource `config` returned a code 4.01 (Unauthorized), meaning that some authorization measure is in place. Overall, we observed 841 2.05 (Content), four 4.01, 105,673 4.04 (Not Found), and 24 4.05 (Method Not Allowed) return codes.

One backend exposes all its sensors, actuators, and values/states via a GET request to `.well-known/core`. To avoid negatively impacting the backend, we have not further studied these exposed resources. However, motivated attackers do not restrain themselves and can misuse the exposed resources to gather intelligence and possibly perform remote actions. For example, for a smart building, this setting could allow for understanding if it is occupied via its sensors and opening doors via its actuators.

#### 6.4.2 *Denial of Service.*

We find 25,928 CoAP backends vulnerable to ZYXEL_000 (30.16%) and 25 vulnerable to CVE-2019-9750 that can be abused to launch amplification attacks. We record the AF for all backends, that is, how much larger the response size is compared to the request. We only consider AFs ≥100%. For ZYXEL_000, we find a maximum AF of 849.06% with an average of 240.61% ($\sigma$=25.80). For CVE-2019-9750, we find the same maximum value but a higher average of 517.43%, at a larger standard deviation ($\sigma$=360.07), signifying that there is a difference in AF across vulnerable backends. The AF for ZYXEL is mainly between 200% and 400%, while the distribution is more sparse for CVE-2019-9750. AFs for other protocols, like the Network Time Protocol (NTP), can be higher, up to 200 times the request size. However, CoAP is a protocol for resource-constrained environments, in which even a lower factor can cause DoS. Finally, we found 212 backends (0.25%) vulnerable to CVE-2018-12679. If exploited, such vulnerabilities could cause the collapse of the entire backend by a DoS attack, thus making all operations and resources unavailable.

**Recommendations.** We suggest developers adopt ACLs to limit access to resources on CoAP backends following the principle of least privilege, i.e., a user should only have access to the resources they need to operate. Further, we deem it important to prevent access to the `.well-known/core` resource, as this would reveal the structure of the backend. Communication should be encrypted when possible (e.g., if connected clients support it). This can be achieved by adopting DTLS. We also encourage updating and applying patches for old vulnerable library versions, as those allow known attacks, such as DoS.

## 6.5 XMPP

**Approach.** We use *nmap* (v7.80) to gather information about the XMPP backends, including their name, version, authentication mechanisms (e.g., PLAIN, DIGEST-MD5), and TLS support [48]. We also employ the *XMPP Compliance Tester* [34] to try registering the account `user` and to test the backend for a set of compliance requirements, like TLS encryption ciphers etc. [110].

**Results.** We successfully connected to 125 out of 702 (17.81%) XMPP backends, which we consider the baseline, while 136 (19.37%) backends were unresponsive (filtered/closed port exceptions).

#### 6.5.1 *Weak Authentication.*

PLAIN is the most common authentication mechanism (56 backends, 44.80%). According to the AUTH PLAIN specifications, username and password are sent from the client to the backend as a base64-encoded string. Sending credentials in this way, that is, without encryption, is a security threat. Malicious actors passively listening to the communication can decode the credentials and use them to log in. PLAIN, being the top authentication mechanism, shows a widespread insecurity of XMPP backends. We also discover DIGEST-MD5 (36 backends, 28.80%) and CRAM-MD5 (33 backends, 26.40%) as authentication methods. Both methods adopt a client-server challenge-response authentication mechanism. In their response, credentials are hashed using the password as the secret key. Generally, DIGEST-MD5 is more secure than CRAM-MD5 as it prevents chosen plaintext attacks [39]. However, considering current computational capabilities, MD5 collisions are becoming achievable, rendering MD5-based authentication methods obsolete and insecure. Finally, we find nine backends (7.20%) with ANONYMOUS authentication, allowing unauthenticated users to access the backend's content.

With *XMPP Compliance Tester*, we can register a dummy user (with username `user`) with no authentication for four backends

(3.20%). Registering a user without requesting a password indicates an insecure implementation of XMPP backends. A malicious actor can send and receive messages without proper authorization, undermining the system's confidentiality and integrity.

#### 6.5.2 *Information Leakage.*

**Demographics.** The main languages of XMPP backends are Russian for 68 (54.40%) backends and English for 30 (24.00%) backends, providing some insight into their geographical distribution.

**Advertised Features.** Among other features, TLS (see Section 6.6 for a detailed investigation into TLS) occurs most often (64 backends, 51.20%). Following TLS are *Roster Versioning* (29 backends, 23.20%) [91] and *In-Band Registration* (24 backends, 19.20%) [90]. Rosters are users' contact lists on the backend. *Roster Versioning* saves bandwidth by not sending unmodified rosters.

**Recommendations.** We encourage developers to drop support for weak and broken authentication mechanisms, like PLAIN, as they undermine the integrity and confidentiality of XMPP backends, potentially allowing attackers to impersonate legitimate users or register new accounts. Moreover, we recommend that operators prohibit user registration without a password.

## 6.6 TLS Adoption for MQTT and XMPP

The IoT ecosystem is highly heterogeneous in systems, devices, and underlying topologies. No standard security solution can be readily applied to all of them. Each scenario requires careful adaption. Nevertheless, TLS provides a fundamental security building block that is widely adopted and supported [13, 77]. Unfortunately, TLS adoption in the IoT domain has been limited because of power/energy concerns [61]. Paracha et al. [77] also found that many IoT devices have wrong TLS configurations, making them vulnerable to attacks. Following, we scrutinize the corresponding backends. We analyze TLS adoption across our TCP-based protocols, MQTT and XMPP. We leverage *testssl* (v3.2rc2) [109] to analyze TLS support and whether cryptographic flaws exist. We extract the supported protocol versions and test for 17 vulnerabilities (e.g., Logjam [64], BEAST [63], SWEET-32 [66]). Additionally, we check whether the certificate has expired or if it was revoked, i.e., if it is in Certificate Authorities (CAs) Certificate Revocation Lists (CRLs).

We test all XMPP backends and sample a random subset of 100,000 MQTT backends (around 25%), which provides statistically significant insight. We automate and parallelize our analysis, using a first timeout of two minutes per backend. We perform our analysis between December 2022 and January 2023.

Table 2 summarizes our results. We successfully connected to 54,503 MQTT and 497 XMPP backends via TLS. Dahlmanns et al. [21] also observed a low TLS adoption rate in their analysis (around 6%). We find a worryingly low fraction of MQTT backends adopting TLS (0.13%) in our dataset, especially considering how critical some of the exchanged messages are. Still far from Dahlmanns et al. results, 2.61% of XMPP backends adopt TLS.

**Protocol Versions and Supported Ciphers.** We find a great share of backends adopting outdated protocol versions, with only 54.65% supporting the latest standard (i.e., only 47 backends support TLS v1.3). Among the TLS-enabled backends, we find that 68.60% support TLS v1 and 70.93% TLS v1.1. The Internet Engineering Task Force (IETF) deprecated both versions in June 2018 based on the severity of discovered cryptographic attacks. Correspondingly, we test whether backends exhibit weaknesses that could be used to mount attacks. We find that 57 backends (66.28%) are vulnerable to BEAST, which affects TLS versions ≤1 and allows attackers to capture and decrypt sessions, rendering encryption useless. Additionally, 40 backends (46.51%) are vulnerable to SWEET-32, a weakness in block ciphers discovered in 2016. Although the vulnerabilities are potentially exploitable, some pre-conditions must be met. For SWEET-32, the exploitability depends on whether the affected ciphers are indeed chosen (proposed by the client, picked by the server), and it generally requires a large number of payloads, the threshold of which may or may not be realistic for an IoT device (depending on connection lengths etc.). Finally, we find 15 backends (17.44%) potentially vulnerable to Logjam, a flaw affecting systems adopting the Diffie-Hellman key exchange with the same prime number, first discovered in 2015. Since then, a 2048-bit shared prime number is considered required.

We additionally analyze the cipher suites the TLS-enabled backends support and report the three most common ones per TLS version (see Table 3). Most of them are not recommended by the Internet Assigned Numbers Authority (IANA), that is, they have not been through the consensus process or have limited scope. Some backends also adopt known weak (broken) cryptographic protocols and hash functions. Specifically, two backends adopt RC4 (2.33%), 40 adopt 3DES (46.51%), 75 adopt SHA-1 (87.21%), and two adopt MD5 (2.33%). These algorithms and hash functions have (long) been deprecated because they are vulnerable to attacks.

**Table 2: TLS-enabled Backends.** We report the number of TLS-enabled backends together with the adopted protocol versions (from oldest to most recent) and their vulnerability to attacks.

| | | MQTT | XMPP | Total |
|---|---|---|---|---|
| **TLS support** | Number | 73 | 13 | 86 |
| | Fraction | 0.13% | 2.62% | 0.16% |
| **TLS version** | Version 1 | 52 | 7 | 59 |
| | Version 1.1 | 54 | 7 | 61 |
| | Version 1.2 | 73 | 13 | 86 |
| | Version 1.3 | 42 | 5 | 47 |
| **Vulnerabilities** | BEAST | 50 | 7 | 57 |
| | SWEET-32 | 37 | 3 | 40 |
| | Logjam | 12 | 3 | 15 |

**Table 3: Supported Ciphers by Protocol Version.** Top three most adopted cipher suites by TLS backends for the analyzed TLS versions. We mark the protocols not recommended by IANA (✖).

| | Cipher Suite | | No. |
|---|---|---|---|
| **v1** | RSA_WITH_AES_256_CBC_SHA | ✖ | 58 |
| | ECDHE_RSA_WITH_AES_128_CBC_SHA | ✖ | 58 |
| | RSA_WITH_AES_128_CBC_SHA | ✖ | 58 |
| **v1.1** | RSA_WITH_AES_256_CBC_SHA | ✖ | 60 |
| | RSA_WITH_AES_128_CBC_SHA | ✖ | 60 |
| | ECDHE_RSA_WITH_AES_256_CBC_SHA | ✖ | 59 |
| **v1.2** | ECDHE_RSA_WITH_AES_256_GCM_SHA384 | | 83 |
| | RSA_WITH_AES_128_GCM_SHA256 | ✖ | 83 |
| | ECDHE_RSA_WITH_AES_128_GCM_SHA256 | | 73 |
| **v1.3** | AES_256_GCM_SHA384 | | 47 |
| | AES_128_GCM_SHA256 | | 47 |
| | CHACHA20_POLY1305_SHA256 | | 45 |

**Certificates.** We find two expired certificates. On average, certificates have an expiration date of ~200 days with one extreme outlier of 982 years. When available, we also retrieved the CRLs from the CA for each backend's certificate and determined if the backend's certificate was revoked. We found no revoked certificate.

Additionally, we analyzed whether the hostname and Common Name (CN) or Subject Alternative Names (SANs) contained in the certificates match. We find 26 of 85 certificates mismatch (30.59%). This suggests that IoT devices do not properly validate TLS certificates, rendering them susceptible to MITM attacks, or that they use certificate pinning with its associated problems. This is important as devices might contain old certificates that are replaced on the backend, but the devices cannot reach the backend. This mismatch "bricks" the devices or exposes them to security issues: Devices can no longer recognize valid backends and cannot download new certificates or security patches [41].

## 6.7 Longitudinal Analysis

We performed our initial measurements between August and November 2022. Here, we investigate on how the availability of backends changed over time and how their security posture evolved in September/October 2023 and January 2024.

**IoT Protocols Analysis.** We first repeat our security assessment between September 15th–30th, 2023. We study 29,077 MQTT, 28,974 CoAP, and 124 XMPP vulnerable backends. Overall, 13,257 MQTT (45.59%), 13,573 CoAP (46.85%), and 35 XMPP (28.23%) backends are now unresponsive or offline. IP addresses can be volatile, and backends may no longer be reachable at the same address. We overcome this shortcoming by scrutinizing the domains we collected and analyzing the backends' new IP addresses, focusing on 17,742 MQTT, 14,288 CoAP, and 48 XMPP vulnerable backends. In some cases, Shodan does not report any associated domain with the IPs, leading to fewer domains.

Some responsive backends no longer exhibit any security vulnerability, indicating that security issues were addressed. We find that 314 MQTT (1.08%), 149 CoAP (0.57%), and six XMPP (6.06%) backends are no longer vulnerable as of September 2023. Unfortunately, other backends have worse security. While 14 CoAP backends (-6.60%) are no longer vulnerable to CVE-2018-12679, 23 new backends (+10.85%) are now vulnerable. Similarly, 147 more MQTT backends (+35.85%) turned vulnerable to DoS threats, while only 84 backends (-20.49%) addressed the vulnerabilities. On a positive note, 185 CoAP backends (-23.75%) exposed fewer resources than in our previous analysis (16, +2.05%, exposed more resources). Finally, 262 MQTT backends (-2.17%) suffer from fewer CVEs from adopting older library versions. This is reflected in the number of backends that use newer updated library versions: 527 MQTT backends use newer software. Interestingly, we also discovered 40 backends that went backward to an older and vulnerable version.

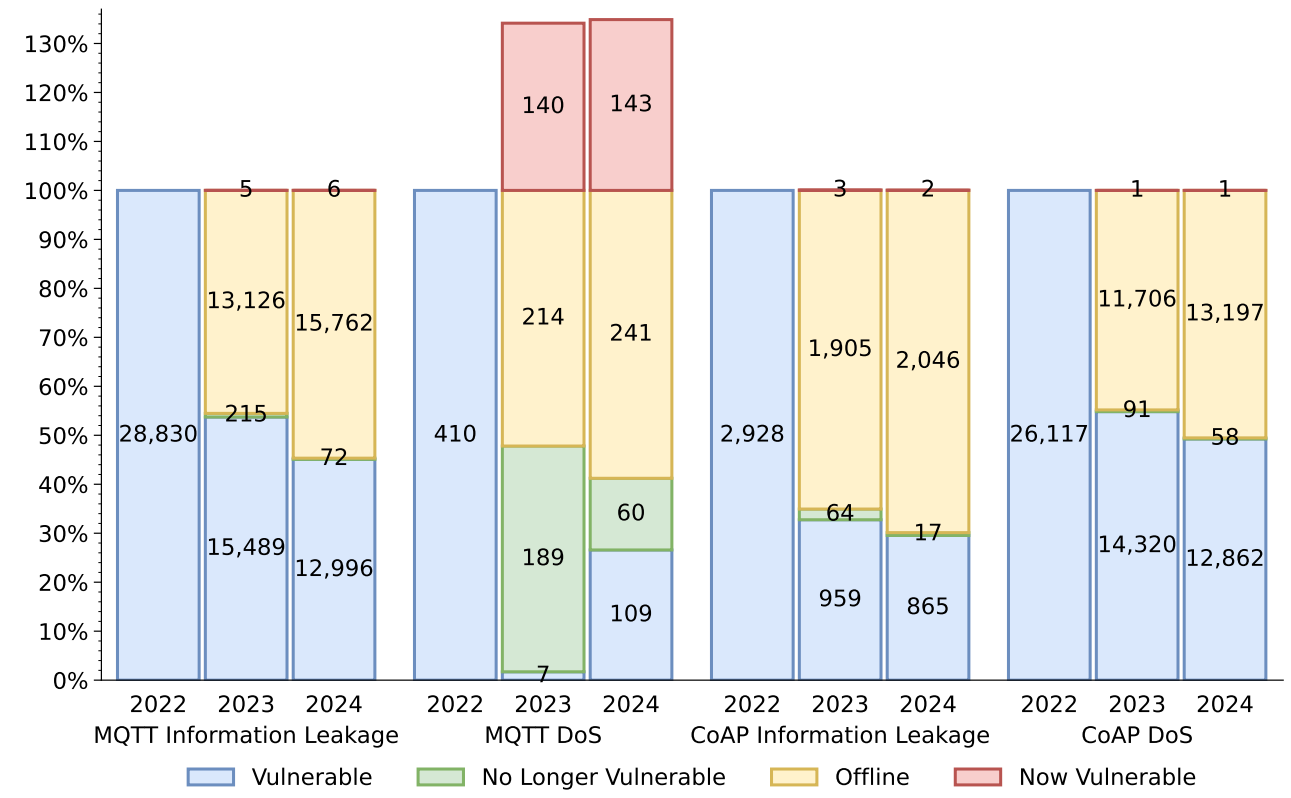

**Figure 4: Vulnerable MQTT and CoAP Backends over the Years.** We omit XMPP backends from the figure for clarity due to their low numbers. We clearly see the volatility of IoT datasets: After just one year, almost half of the backends are no longer reachable. Overall, more backends have fixed the vulnerabilities over the years. Interestingly, for MQTT backends, the number of newly vulnerable backends for DoS threats also greatly increases, signaling that, in such cases, older software versions are being rolled out.

We repeat our analysis a second time between January 23rd–31st, 2024, after we sent our disclosure emails, to understand whether developers who had been made aware of the vulnerabilities addressed them (see Appendix B for an in-depth discussion of our disclosure process). Overall, we encounter a similar instability as in our second scan and find 15,909 MQTT (54.71%, +12.12%), 15,203 CoAP (52.47%, +5.62%), and 38 XMPP (30.65%, +2.42%) backends unresponsive or offline, showing how, over time, datasets age quickly, resulting in 13,168 MQTT, 13,771 CoAP, and 86 XMPP responsive backends.

In addition to the backends that fixed their vulnerabilities in 2023, 74 MQTT (0.56%) and 72 CoAP (0.52%) backends no longer suffer from vulnerabilities. Similar to 2023, 145 MQTT (+30.66%) backends became more vulnerable to DoS attacks, while only 127 (-26.85%) addressed the issue. For the remaining vulnerabilities, we observe slight improvements. We report the overall trends for MQTT and CoAP backends in Figure 4. We do not show MQTT backends that suffer from weak authentication, i.e., the backends with credentials-based authentication, because they have not changed over the years besides the number of backends that became unreachable.

**TLS Analysis.** We repeat our TLS analysis between October 15th–30th, 2023. Given reachability instability for IP-based backends, we successfully analyzed 38,034 MQTT and 48 XMPP backends, of which only 25 MQTT (0.07%) and 3 XMPP (7.89%) backends support TLS. Interestingly, the two backends that previously served expired certificates provide the same old certificates, signaling poor security practices. We find 4 MQTT backends that show worse security in their TLS configuration. One now supports the outdated SSLv3 version, making it vulnerable to more attacks (e.g., SWEET-32).

# 7 Discussion

Following our study of the individual protocols and TLS, we discuss characteristics and statistics of our results, contextualize findings, and provide details on trends.

**Results per Threat Class.** We summarize our analysis results in Table 4. Considering the number of vulnerable backends per category, we see that a large fraction of reachable MQTT (11.47%) and XMPP (72.95%) backends are vulnerable to information leakage threats, with topic enumeration being a superset of backend fingerprinting for MQTT. For all reachable MQTT backends for which we collected topic names, we obtained fingerprint information, such as the software version or the number of clients. CoAP backends are particularly vulnerable to DoS attacks or are even enabling them (30.38%). This may be due to CoAP being UDP-based. Noteworthy is the large amount of CoAP backends that act as amplifying reflectors (30.18%), posing a severe risk to the Internet.

**Results per Geographical Location.** Concerning MQTT topic enumeration by country, most information leakage occurs for backends located in China (729,425), followed by the US (422,771) and Germany (247,663). Except for China, this trend does not reflect the geographical distribution of brokers we observed in Section 5.1. On average, these countries reveal more topics per backend than the countries with most MQTT brokers.

We observe that 24,519 (94.57%) CoAP backends that are vulnerable to ZYXEL_000 are located in Russia, and they represent the vast majority of Russian CoAP backends (99.68%). The vulnerability ZYXEL_000 affects various Keenetic routers and enables DDoS attacks, potentially rendering thousands of devices unavailable. One reason for this localization in Russia could be an ISP providing vulnerable routers to its customers. Albeit Russian CoAP backends account for the majority of DoS amplifiers and reflectors, they expose fewer resources, with only 0.21% backends exposing any resources. Somewhat counterintuitively, backends in Europe and the US exhibited more information leakage in 2022, with 30.98% and 39.62% backends, respectively, exposing resources. This is despite stricter privacy regulations, like the European General Data Protection Regulation (GDPR). Interestingly, this improved in 2023, with now only 20.77% European and 27.99% US CoAP backends exposing resources. One reason might be that, in addition to the GDPR, the California Privacy Rights Act (CPRA) [14] came into effect on January 1st, 2023, requiring companies to put more care into handling users' data.

XMPP backends in the EU are generally more secure than those in the US, China, or Russia: 25.58% EU backends are vulnerable compared to 44.44–56.25% in other regions.

Overall, in 2023, compared to our 2022 results, the percentages of vulnerable backends in the analyzed countries slightly decreased for all vulnerability categories except for DoS threats for MQTT, for which we instead witnessed a slight increase in vulnerable backends in Europe (21 to 30), the US (18 to 22), and China (259 to 294).

**Results per Cloud Providers.** Taking a look at backends' deployments, we observe that MQTT backends hosted on larger cloud providers (AWS, Google, Azure, Alibaba) exhibit worse security posture than those hosted on *Other* (see Table 5). While we cannot identify a clear trend for XMPP, it unambiguously reverses for CoAP: Cloud providers generally show better security.

The trend of generally better security is, however, not uniform. Cloud-hosted CoAP backends leak more information than *Other*. In 2022, over one-third (34.76%) of AWS-hosted backends exposes at least one resource, contrary to only 0.68% *Other* backends. This improves substantially in 2023 for cloud-hosted backends: Only 17.63% AWS-hosted backends leak resources and the number of Google-hosted backends that expose resources halves (50.77% to 24.61%), indicating that security is receiving some attention. Fortunately, considering their network capacity, almost no cloud-hosted backends are vulnerable to DoS amplification vulnerabilities (0.75%), while almost one-third of *Other* backends are (30.58%). This number remains stable in 2023, suggesting that these providers adopt the latest security updates to mitigate abuse while other operators do not. At the same time, backend operators are still responsible for configuring the cloud-hosted backends properly to prevent information leakage, and the complementary disparity we observed provides a unique opportunity for future human factors research.

**Table 4: Vulnerable Backends per Threat Class.** We group the results of our individual analyses by vulnerability to provide a more comprehensive overview. A considerable share of backends in our dataset is vulnerable, potentially affecting the confidentiality, integrity and availability of user data.

| | Threat Class & Analysis | # Vulnerable | Fraction |
|---|---|---|---|
| MQTT | *Information Leakage* | 28,830 | 11.47% |
| | Backend Fingerprint | 23,120 | 9.20% |
| | Topic Enumeration | 28,830 | 11.47% |
| | *Weak Authentication* | 12,071 | 4.80% |
| | *Denial of Service* | 410 | 0.16% |
| | Known Vulnerabilities (backend) | 410 | 0.16% |
| CoAP | *Information Leakage* | 2,928 | 3.41% |
| | Backend Fingerprint | 2,864 | 3.33% |
| | Resource Listing | 779 | 0.91% |
| | *Denial of Service* | 26,117 | 30.38% |
| | Amplification Factor (client) | 25,939 | 30.18% |
| | Known Vulnerabilities (backend) | 212 | 0.25% |
| XMPP | *Information Leakage* | 89 | 72.95% |
| | Backend Fingerprint | 89 | 72.95% |
| | *Weak Authentication* | 59 | 48.36% |
| | Supported Authentication Mechanism | 56 | 45.90% |
| | Compliance | 4 | 3.28% |

**Table 5: Vulnerable Backends per Provider.** Number of vulnerable backends by provider and protocol together with the respective fraction (computed on the total backends belonging to a specific provider). Considering their low numbers, we merge Cloudflare and Oracle with *Other*.

| | | AWS | Google | Azure | Alibaba | Other |
|---|---|---|---|---|---|---|
| MQTT | Total | 4,036 | 606 | 854 | 131 | 245,753 |
| | Vulnerable | 2,883 | 409 | 625 | 45 | 25,091 |
| | Fraction | 71.43% | 67.49% | 73.18% | 34.35% | 10.21% |
| CoAP | Total | 397 | 65 | 100 | 6 | 85,389 |
| | Vulnerable | 3 | 0 | 0 | 0 | 26,114 |
| | Fraction | 0.76% | 0.00% | 0.00% | 0.00% | 30.36% |
| XMPP | Total | 7 | - | 1 | - | 118 |
| | Vulnerable | 5 | - | 1 | - | 93 |
| | Fraction | 71.43% | - | 100.00% | - | 78.81% |

**Table 6: Count of Responses.** We grouped the responses we received from the disclosure emails we sent into categories.

| Type | Count |
|---|---|
| "Listed CVEs do not apply. Provide more information." | 23 |
| "We have informed the responsible parties." | 15 |
| "We fixed the issues with your information." | 11 |
| "We do not have time." | 3 |
| "We will let you know what we will do." | 2 |
| "The vulnerable client has been blocked." | 1 |
| Automatic Reply | 428 |
| Failed Delivery | 700 |

**Self-hosted AWS vs. Managed AWS.** Matching hostnames in our dataset against regexes of hostnames of services managed by AWS [86], we find only 125 instances that are managed MQTT backends (3.10% of AWS backends). Interestingly, we failed to connect to all 125 backends, possibly because they were unavailable or implement ACLs properly. Indeed, AWS IoT adopts certificates to authenticate clients, which may prevent us from successfully connecting. They might also use Amazon Cognito to obtain (temporary) limited-privilege credentials. However, this does imply that these backends are secure. Companion apps might use credentials to authenticate their connection to the backends and carelessly hardcode the credentials in the app code [45]. Since we do not know if backends are associated with any app, we leave managed AWS backends for future work and consider them secure.

Looking at Table 5, we can see that around 71.43% self-hosted AWS backends are vulnerable to some threat class, from information leakage to DoS. Hence, we highlight the risk of inexperienced users misconfiguring AWS instances and potentially exposing sensitive information; this is less likely in the case of instances created directly by AWS, as our results show.

**Ethical Considerations.** Active measurements, like ours, raise ethical concerns that demand proper consideration. For our evaluation, we followed guidelines defined by the Ethics Review Board (ERB) of the University of Twente, which reviewed and approved our study (see Appendix A for an in-depth discussion). When devising our measurement methodology, we put particular care into performing analyses that do not alter state. We do not control the backends and interfering with their operation, such as impeding or disrupting their service, would clearly raise ethical concerns. Therefore, we do NOT perform any actions that could compromise the correct functioning of the backends, such as ones leading to DoS. Further, we only provide aggregated data that cannot be associated with any specific service.

**Vulnerability Disclosure.** We have started a Coordinated Vulnerability Disclosure (CVD) process to inform developers and operators about the issues we discovered, which is still ongoing. We report for each backend, the scan date, our methodology, the vulnerabilities we found, and suggestions on how to address them. We have sent 2,132 emails (for 15,820 backends) and received 1,173 responses as of January 2024, categorized in Table 6. We provide an in-depth discussion in Appendix B, as well document our experience and still open challenges with vulnerability disclosure at this scale in an auxiliary workshop paper [16].

## 8 Limitations & Future Work

Our results clearly show a problematic security immaturity in the IoT ecosystem. Concurrently to our work, Yaben et al. [112] identified over 1M endpoints using ZMap in December 2023, and painted a similar bleak picture on "abandoned or neglected" servers based on unpatched/outdated software and misconfigurations, such as weak authentication and expired certificates. We leave a comparison for future work. Following, we discuss possible threats to the validity of our study and further challenges for future work.

First, we acknowledge that our dataset might intrinsically contain geographical or provider bias. We could not find many backends in prior IoT traffic and were unsuccessful when asking other researchers to share theirs. At the same time, the high number of backends on Shodan shows a problematic lack of coverage of existing datasets and limited visibility into the ecosystem. Further, our dataset is mostly IP-based, as not all Shodan results include a hostname. As discussed in Section 6.7, relying on IPs can hinder stability. In future, we aim to gather a more heterogeneous dataset.

Second, we were limited in the range of vulnerability analyses we could perform. To preserve the correct functioning of the analyzed services and avoid disruptions or interruptions, we did not perform invasive measurements, for example, testing carefully crafted malformed payloads. Our analysis cannot guarantee complete insights about the security posture of IoT backends because there might be other potential threats they are exposed to, which we did not investigate. Nevertheless, our results clearly show that additional steps must be taken to improve IoT security.

Further, we cannot rule out that some backends we analyze are not part of the IoT. Some might be IoT-related or IoT-adjacent. Performing the necessary experiments to accurately assess whether a backend speaking an IoT-focused protocol is truly IoT would cross ethical and likely legal boundaries. Specifically, if we wanted to investigate the backends further and test if the connected clients are IoT devices, we would need to perform invasive measurements of the connected clients and instruct them to perform some action that we can use to determine that they are IoT. This is clearly much more ethically and legally challenging, if not downright impossible.

We also cannot rule out that our dataset contains honeypots. To the best of our knowledge, only one IoT-focused honeypot exists currently [103], which provides only basic functionality and has not been deployed widely. To quantify this issue, we employ Shodan Honeyscore [97], which has been integrated into the regular crawlers since its first release. We find only 36 instances of the MQTT backends for which the "honey*" keywords have been set, indicating that they almost certainly represent negligible noise. Unfortunately, without more invasive measurements, we cannot rule out that other backends are honeypots.

Finally, we focused on three widely adopted IoT protocols and considered more general protocols, like HTTP, out of scope. Despite HTTP usage in the IoT ecosystem [26], distinguishing IoT HTTP backends from non-IoT ones is extremely challenging. It requires a semantic understanding of its API and manual inspection, which does not scale and requires invasive analyses. We leave their study for future work. Other non-application-layer IoT protocols, like Z-Wave, Zigbee, or RFID, have been studied by prior work [29, 114] and are local only, while we analyze public remote backends.

## 9 Related Work

**IoT Protocol Security.** Maggi et al. [54] investigated the security of MQTT and CoAP. They found sensitive healthcare data exposed by insecure MQTT brokers, such as patients' PII and ambulance locations. Additionally, they found 365,000 CoAP backends exposing network credentials. Palmieri et al. [76] showed the insecurity of MQTT backends, with 24,361 backends (60.38%) allowing clients to simply connect. They proposed MQTT-SA, a tool to assess MQTT deployments' security and detect possible misconfigurations. Jia et al. [43] successfully exploited MQTT device sharing or access revocation weaknesses to send unauthorized messages using "Will and Retained." Andy et al. [7] also investigated the implementation issues of MQTT, such as lack of authentication and encryption. Paracha et al. [77] studied how different IoT devices use TLS by collecting device traffic for two years. Their results show that some devices adopt old or insecure protocol versions or lack certificate validation. However, the studies and methods of prior work do not scale and are not applicable for publicly exposed backends, as they are invasive and potentially cause crashes.

**IoT Analysis at Scale.** Saidi et al. [86] studied the geographic location of IoT backend providers. They found that ~35% of IoT traffic at a major European ISP is going to providers located outside of Europe, raising regulatory concerns. Srinivasa et al. [102] perform Internet-wide scans on six protocols, including those we study in this paper. They find over 1.8 million misconfigured IoT devices that can either be infected with bots or be leveraged for a (D)DoS amplification attack. Dahlmanns et al. [21] studied TLS adoption of ten Industrial IoT protocols, showing a low deployment rate (6.5%) and other widespread security issues (e.g., outdated protocol versions). Other work focused on large-scale identification of IoT devices and related events based on network traffic characteristics and packet signatures [25, 57, 74, 80, 99, 108]. Recent work investigated the IoT ecosystem at scale by focusing on mobile companion apps [15, 23, 82, 84] looking for security and privacy issues of devices without direct access to them. For example, most recently, IoTFlow [94] statically reconstructed network-related data such as URLs, including backends, contacted by 9,889 companion apps. However, except concurrent work by Yaben et al. [112], prior work did not study IoT backend security at scale.

## 10 Key Takeaways

**Security Immaturity of IoT Backends.** Prior work on studying the security of backend for IoT protocols lacks coverage. We fill this gap by gathering an extensive dataset of over 337,464 active backends (251,382 MQTT, 85,957 CoAP, and 125 XMPP). One in six backends (17.24%, 58,175 backends) suffer from Weak Authentication, Information Leakage, or DoS.

**Weak Authentication.** 59 XMPP backends (48.36%) adopt weak authentication mechanisms, like PLAIN, and 12,071 MQTT backends (4.80%) use credentials-based authentication in plaintext over the network, enabling attackers to easily exfiltrate data.

**Information Leakage of Sensitive Data.** 31,847 MQTT, CoAP, and XMPP backends (9.44%) expose data ranging from version information to topic/resource names to messages for up to 40,017 security-sensitive MQTT topics.

**Denial of Service.** Nearly one third of CoAP backends (25,939; 30.18%) enable DoS amplification attacks. Although patches for the underlying vulnerabilities are already available, most backends use outdated vulnerable software, enabling DoS attacks.

**An Absence of Transport Layer Security (TLS).** Today, a negligible 0.16% of the TCP-based backends use TLS, and the majority of them use old versions (70.93%, version < 1.1) and suffer from known vulnerabilities (66.28% BEAST, 46.51% SWEET-32). Adapting and adopting TLS in resource-constrained devices could prove a simple approach to quickly improve security.

**Cloud Hosting.** The majority of cloud-hosted backends (30,819; 96.96%) are run by AWS, Google, or Azure, supporting their dominance. However, utilizing a large cloud provider does not automatically lead to better security, for example, cloud-hosted MQTT backends exhibit worse security.

**Large-scale and Highly Problematic.** The identified issues affect non-critical and critical infrastructure alike: an MQTT backend, likely related to a power facility, exposes 36,254 topic names, while 99.68% of Russian CoAP backends enable amplification attacks due to the widespread adoption of Zyxel routers.

## 11 Conclusions

In this paper, we perform a large-scale measurement of the security posture of over 337,000 IoT backends that use MQTT, CoAP, or XMPP, focusing on three main threats: information leakage, weak authentication, and DoS potential. We find that many deployments for all three protocols are vulnerable: 31,847 of the reachable backends (9.44%) expose (sensitive) information, a conspicuous fraction of CoAP backends (30.18%, 25,939 backends) are vulnerable to amplification attacks, and only a negligible number of MQTT and XMPP backends adopt TLS (0.16%), of which 70.93% use outdated protocol versions (< 1.1). Our study provides evidence for a troubling immaturity of security in the IoT ecosystem, which was not analyzed thoroughly at scale before. We responsibly disclosed the identified issues to the affected parties, support their remediation efforts, and hope to improve their security awareness.

## Acknowledgements

We thank our reviewers for their valuable comments and inputs to improve our paper. We also thank Ting-Han Chen and Jeroen van der Ham for their assistance in the disclosure process.

This work is based on research supported by the Vienna Science and Technology Fund (WWTF) and the City of Vienna [Grant ID: 10.47379/ICT19056], the Deutsche Forschungsgemeinschaft (DFG, German Research Foundation) under Germany's Excellence Strategy - EXC 2092 CASA - 390781972, the INTERSECT project, Grant No. NWA 1160.18.301, funded by the Netherlands Organisation for Scientific Research (NWO), the Internet Society Foundation, and SBA Research (SBA-K1), a COMET Centre within the framework of COMET – Competence Centers for Excellent Technologies Programme and funded by BMK, BMDW, and the federal state of Vienna. The COMET Programme is managed by FFG. Any opinions, findings, and conclusions or recommendations expressed in this material are those of the authors and do not necessarily reflect the views of the respective funding agencies.

## A Ethical Considerations

We performed large-scale active measurements, also called scans, of real-world deployments. This prompts important ethical considerations, similarly to prior studies [18, 81, 101]. We followed the guidelines and best practices established in the Menlo report [8] and also discussed in recent work on cybersecurity research and network measurements [53, 78, 79]. Our study has been approved by the Ethics Review Board (ERB) of the University of Twente in the Netherlands, from where we also performed our measurements. The ERB assessed our setup and measurement methodology. In recognition of a historic lack of computer science expertise in the ethics review process, our ERB operationalizes the inclusion of cybersecurity expertise in its review, including guidance on coordinated vulnerability disclosure [83] (see also Appendix B).

**Active Measurement Setup** To prevent potential harms to the scanned backends we put the following precautions in place: First, we limited the number of requests that we perform with our scanners to limit the impact we have on backends. The machine that we used to perform our measurements has a static IP address in our IP address space from the University of Twente and has a clear registered abuse handle. We also set up a reverse DNS entry with a descriptive DNS name for the IP address and we host an informative web page on the same machine (reachable directly via IP address and DNS name). These measures aimed to quickly and clearly inform the backends' developers and maintainers about the nature of our measurement (e.g., by directly seeing `iotscan.eemcs.utwente.nl` in their log files and then visiting the website that provides more details), so that they could understand the scope of our study, and could contact us to request even more details or to be excluded from our study. We did not require a reason to to be excluded from our measurements. Note that we received no exclusion requests.

Second, we only conducted *non-invasive* tests to not alter the state of the analyzed backends, carefully considering the trade-off between the utility of our study and potential harms. We only detect vulnerabilities and we did *not* attempt to exploit them. This limits the accuracy of our findings, as in some cases the identified vulnerabilities may not actually be exploitable. We thoroughly evaluated the trade-offs that could cause unintended harm by our measurements with respect to its benefits, which includes providing valuable knowledge to the scientific community and practitioners to protect users by discovering and reporting critical vulnerabilities, so that they can be addressed. We are convinced that the contributions we make in our work, especially considering the plethora of potential threats that we discovered within the IoT to privacy, self-determination, and even life, considerably outweigh the remaining minimal risks of our measurement.

Third, we did not read or store sensitive data. Albeit data could include sensitive information, such as usernames, we only collect the minimally necessary metadata (i.e., topics and resource names) to perform our assessment. We do not collect any potentially sensitive content, to mitigate potential unknown risks. Thus, our analysis is a lower bound, as other (and more) sensitive data might also be exposed. We test if resources (e.g., `/password`) are exposed via HEAD requests and the response code, that is, an existence check, and we do not request or receive any content.

Finally, we do not disclose the affected backends and put particular care into anonymizing our results by only presenting aggregates. When discussing case studies, we do not provide information that links to the specific backend. We will allow researchers to access our anonymized dataset after verifying their roles and institutions.

## B Vulnerability Disclosure

Given the scale of our study and the many vulnerable backends we identified, responsible disclosure is particularly challenging: It involves the operators of tens of thousands of backends. Therefore, we are collaborating with the Dutch Computer Emergency Response Team (CERT) and Cyber Security Center. We are following the coordinated vulnerability disclosure approach proposed by Reidsma et al. [83] and with the support of Ting-Han Chen and Jeroen van der Ham at the University of Twente.

First, we started our disclosure process with the two motivating examples we discussed in Section 4. We emailed the affected parties via the contact information we found on their respective websites in May 2023. We sent an informative email stating who we are, the scope of our research, the vulnerabilities we found, how they could affect their backend, and suggestions on how to address them. We then sent follow-up emails in the following weeks, the first after 21 days and the second after 60 days. We only received feedback from one of the two impacted apps.

Second, for our large-scale study, we check whether an IP address falls within the IP address ranges of major cloud service providers (AWS, Google Cloud, etc.), in which case we disclose our findings directly to the providers. For the remaining backends, we utilize `WHOIS` data to extract their associated email addresses. We extract this information twice, for our first measurement (October 2022) and for our second one (September 2023), as addresses might have changed [104]. We extracted one or more email addresses for 273,151 MQTT, 290,901 CoAP, and 125 XMPP backends. We started our large-scale disclosures in November 2023 for the vulnerable backends affected by a CVE, accounting for the need to understand the evolution of security and privacy in the IoT ecosystem undisturbed while minimizing risks. Importantly, we have been working with the Dutch Cyber Security Center since May 2022, i.e., well before our first measurement, to determine the best and most effective disclosure approach that minimizes overall harm.

We grouped all backends for which we reconstructed the same email address in a single email to reduce the total number, such as those by cloud providers and other large hosting providers. Some maintainers and developers may require more information and suggested solutions, which is why we drafted a file with an in-depth description of the identified vulnerabilities and how they impact their services. Moreover, although we cannot give detailed information on how to address vulnerabilities as we do not have access to their backends, we provide some general guidelines. For example, when vulnerabilities stem from old library versions, we suggest to update to a newer version. Similarly, for information leakage issues, we advise to:

(1) Adopt authentication measures, e.g., (at least) passwords.
(2) Adopt ACLs to prevent users from reading (all) messages, e.g., so that only admins can read sensitive topics.
(3) Encrypt communication, e.g., using (D)TLS.
(4) If the backend does not have to be exposed to the Internet, protect it behind a firewall (blocking incoming connections from outside the organization).

We refer the interested reader to further details on our experience as of January 2024, including still open challenges with vulnerability disclosure at scale in our workshop paper [16].

## C Example MQTT Messages Leaking PII

To illustrate the sensitive nature of user data backends can expose we investigated two MQTT backends extracted from mobile companion apps in two case studies in Section 4.

Listing 1 shows an example messages an attacker could read from the MQTT backend of a *Heart & Lung Health Monitoring Device*, exposing a users age, gender, location, and health indicators. Similarly, Listing 2 shows an example message obtained from the MQTT backend of a *Smart Car Dongle*, exposing the car's location, engine statistics, and status of the anti-theft alarm.

For both examples we omit any information that could identify the user or the company. The developer of the health monitor has started remediation procedures following our disclosure. The developer of the car dongle has not responded to our disclosure.

```
"topic": "-/-/live-broadcast",
"payload": {
    "gender": "-",
    "age": -1,
    "time_zone_utc_offset":-1,
    "altitude": -1,
    "longitude": -1,
    "latitude": -1,
    "type": "",
    "userName": "User Name",
    "shock": -1,
    "breathingRate": -1,
    "uniqueID": -1,
    "strain": -1,
    "hearRate": -1,
    "cadence": -1,
    "distance": -1,
    "pace": -1,
    "userID": -1
}
```

**Listing 1: Example MQTT Message and Topic that we Observed for the Heart & Lung Health Monitoring Device.**

```
"topic": "-/-",
"payload": {
    "speed": -1,
    "rpm": -1,
    "coordinates": [-1, -1],
    "distance": -1,
    "coolant": -1,
    "voltage": -1,
    "trip_time": -1,
    "status": "-",
    "fuel_consumption_1": -1,
    "fuel_consumption_2": -1,
    "gps_satellite_count": -1,
    "gsm_signal_quality": -1,
    "load": -1,
    "IMAP": -1,
    "IAT": -1,
    "air_flow": -1,
    "long_term_fuel_trim": -1,
    "absolute_throttle_position": -1,
    "fuel_ratio_coefficient": -1,
    "direction": -1,
    "id": "-"
}
```

**Listing 2: Example MQTT Message and Topic that we Observed for the Smart Car Dongle.**

## D MQTT Connected Clients

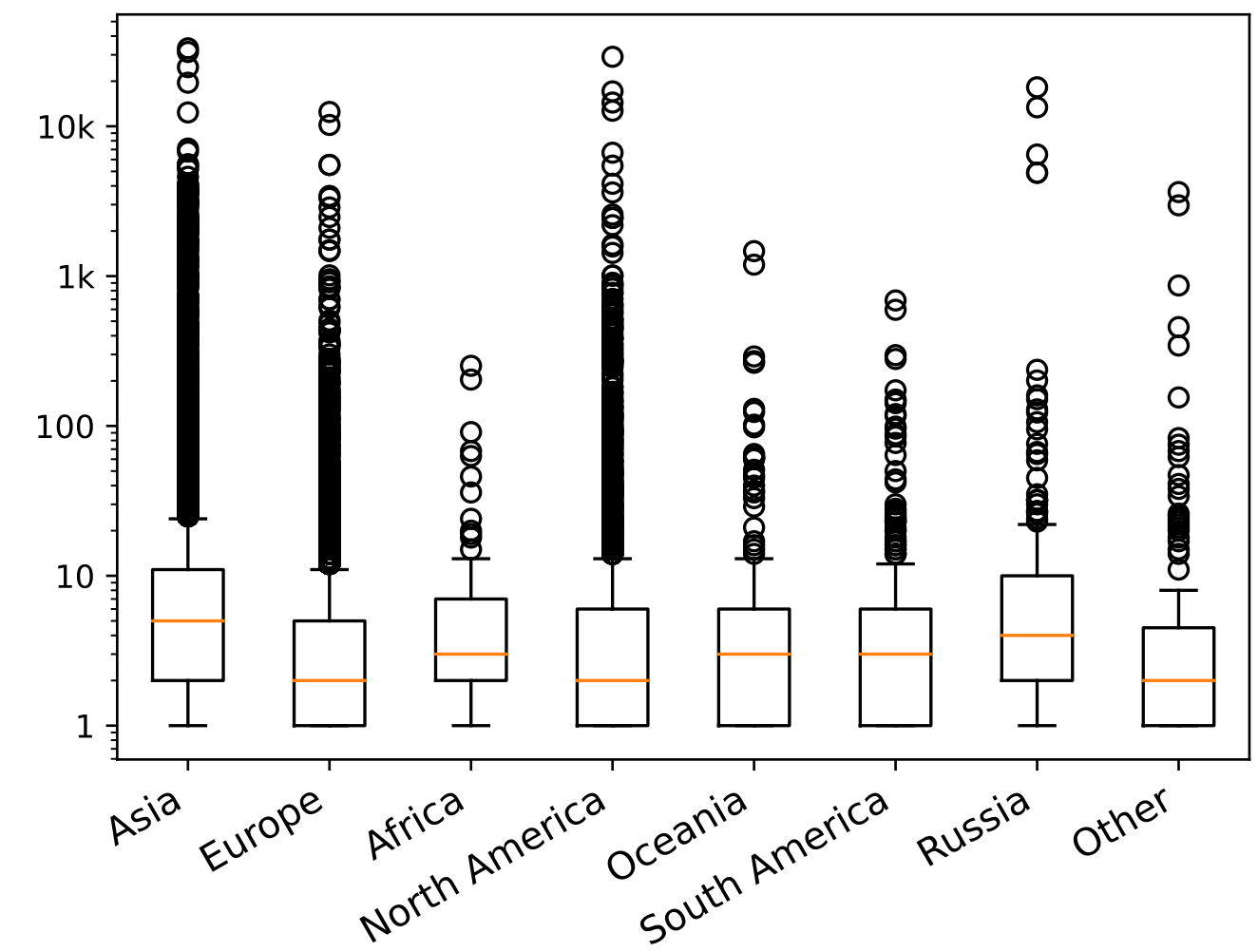

**Figure 5: Number of Connected Clients to Individual MQTT Backends Grouped by Continents.** We see how Asia has the most dense tail, indicating that a great share of MQTT backends are located in that geographical area. Conversely, the densities of Africa, Oceania and Russia are lower.

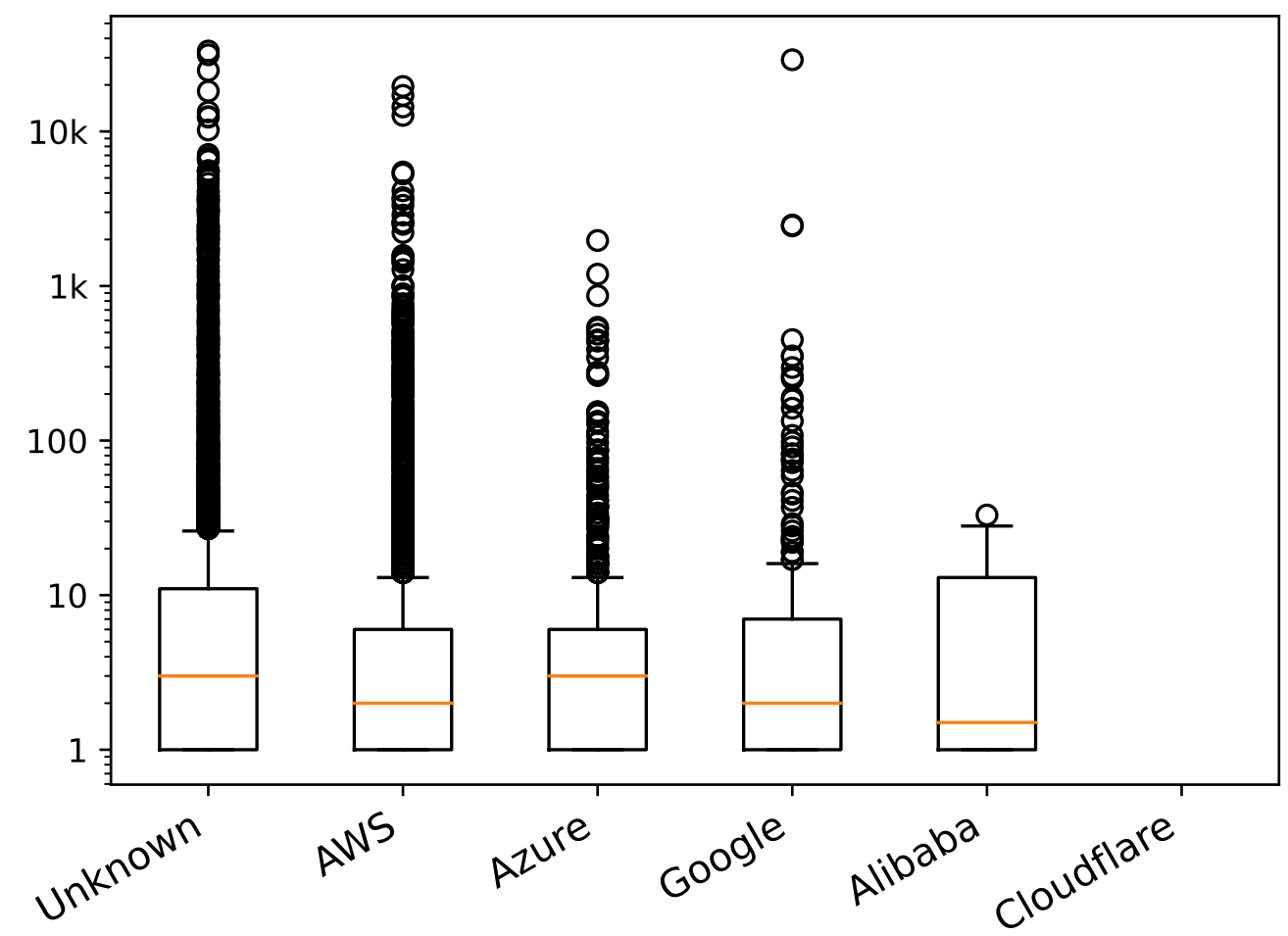

**Figure 6: Number of Connected Clients to Individual MQTT Backends Grouped by Providers.** Alibaba brokers show fewer connection but a trend cannot be inferred because of a low number of observations for Alibaba.